\documentclass{aa}
\usepackage{graphicx}
\def\masyr{mas\,yr$^{-1}$}
\def\kms{km\,s$^{-1}$}
\def\farcm{\hbox{$.\mkern-4mu^\prime$}}
\def\farcs{\hbox{$.\!\!^{\prime\prime}$}}
\def\spose#1{\hbox to 0pt{#1\hss}}
\def\lta{\mathrel{\spose{\lower 3pt\hbox{$\sim$}}
    \raise 2.0pt\hbox{$<$}}}
\def\gta{\mathrel{\spose{\lower 3pt\hbox{$\sim$}}
    \raise 2.0pt\hbox{$>$}}}
\begin{document}
\title{Radial velocities in the globular cluster $\omega$ Centauri}
\titlerunning{Radial velocities in $\omega$ Centauri}
\authorrunning{Reijns et al.}
\author{R.~A.\ Reijns\inst{1},
        P. Seitzer\inst{2},
        R.\ Arnold\inst{1}\thanks{{\it Present address:\/}
        School of Mathematical and Computer Sciences, University of Wellington,
        New Zealand, (Richard.Arnold@mcs.vuw.ac.nz)},
        K.~C.\ Freeman\inst{3},
        T.\ Ingerson\inst{4},\\
        R.~C.~E.\ van den Bosch\inst{1},
        G.\ van de Ven\inst{1},
        and P.\ T. de Zeeuw\inst{1}}
\offprints{P.T.~de Zeeuw}
\institute{
   Sterrewacht Leiden, Postbus 9513, 2300 RA Leiden, The Netherlands
   (omegacentauri@reijns.com, dezeeuw@strw.leidenuniv.nl)
   \and
   Department of Astronomy, University of Michigan, Ann Arbor, MI 48109, USA
   (seitzer@umich.edu)
   \and
   Research School of Astronomy \& Astrophysics, Australian National
   University, Mt.\ Stromlo Observatory, Cotter Road, Weston ACT 2611,
   Australia (kcf@mso.anu.edu.au)\looseness=-2
   \and
   Cerro Tololo Inter-American Observatory, National Optical Astronomy
   Observatories, Casilla 603, La Serena, Chile (tingerson@noao.edu)}
\date{Received 0000 0000, Accepted 0000 0000}
\abstract{We have used the {\tt ARGUS} multi--object spectrometer at
the CTIO 4m Blanco telescope to obtain 2756 radial velocity
measurements for 1966 individual stars in the globular cluster
$\omega$~Centauri brighter than blue photographic magnitude of about
16.5. Of these, 1589 stars are cluster members. A comparison with two
independent radial velocity studies, carried out by Suntzeff \& Kraft
and by Mayor et al., demonstrates that the median error of our
measurements is below 2 \kms\ for the stars brighter than
$B$-magnitude 15, which constitute the bulk of the sample. The
observed velocity dispersion decreases from about 15 \kms\ in the
inner few arcmin to about 6 \kms\ at a radius of 25$'$. The cluster
shows significant rotation, with a maximum amplitude of about 6 \kms\
in the radial zone between 6$'$ and 10$'$. In a companion paper by van
de Ven et al., we correct these radial velocities for the perspective
rotation caused by the space motion of the cluster, and combine them
with the internal proper motions of nearly 8000 cluster members
measured by van Leeuwen et al., to construct a detailed dynamical
model of $\omega$ Centauri and to measure its distance.}

\maketitle
\keywords{radial velocities --
          Galaxy: globular clusters: individual: NGC~5139}

\section{Introduction}
\label{sec:intro}

Globular clusters have been objects of astronomical study for well
over a century (e.g., Pickering \cite{pickering}). They are amongst
the oldest objects in the Galaxy and their stellar content provides
information on star formation and evolution processes, in particular
in the early lifetime of our Galaxy.  Their richness and symmetry, and
the complexity of their dynamical evolution through internal and
external effects, make them very interesting for stellar dynamical
studies (e.g., King \cite{king66}; Spitzer \cite{spitzer}).

Among the globular clusters, $\omega$~Cen (NGC 5139) is particularly
interesting (van Leeuwen, Hughes \& Piotto \cite{vlhp02}). The cluster
was studied long ago for its variable stars (e.g., Bailey
\cite{bailey}; Martin \cite{martin}) and its stellar content (Woolley
\cite{woolley}). It is very massive (Meylan et al.\ \cite{meylan}) and
has an unusually large ellipticity of $\sim$$0.12$. The cluster has a
tidal radius of $\sim$$45'$ (\cite{tkd95}) and is structurally fairly
loose, which makes it possible to study individual stars from the
ground even in the dense central regions. The chemical abundance
distribution is broad, with $-1.8 < \hbox{\rm [Fe/H]} < -0.8$, and
bimodal (Norris \& Da Costa \cite{nordca}, Norris et al.\
\cite{norris96}). Recently, further evidence for multiple stellar
populations was found from colour-magnitude diagrams (Anderson
\cite{and02}; Pancino \cite{pan02}; Bedin et al.\ \cite{bpa+04};
Ferraro et al.\ \cite{fsp+04}; Hughes, Wallerstein \& van Leeuwen
\cite{hwl+04}) and spectroscopy (Piotto et al.\
\cite{pvb+05}).\looseness=-2

Van Leeuwen et al.\ (\cite{llrfz00}, hereafter Paper I) carried out a
proper motion study of 9847 stars in the field of $\omega$~Cen, based
on extensive photographic material obtained with the Yale--Columbia
refractor between 1931 and 1983. Precisions range from $0.10$ \masyr\
at photographic magnitude $13$ to $0.65$ \masyr\ at the limiting
magnitude 16.5. The measurements extend out to 24$'$, i.e., to nearly
half the tidal radius. At a distance of 5 kpc, $0.10$ \masyr\
translates to about 2.5 \kms. A total of 7853 probable cluster members
were identified. The astrometric measurements provide accurate
estimates of two of the three components of the internal motion. They
provide evidence for differential rotation, but only up to an unknown
amount of solid body rotation, which is absorbed in the plate
transformations required to derive the proper motions.  Radial
velocities of similar accuracy or better are needed to obtain the
third component of the internal motion, and to characterize the
rotation of the cluster.\looseness=-2

$\omega$~Cen is an ideal cluster for a radial velocity study because
of its high systemic radial velocity of 232.8 \kms\ (e.g., Meylan et
al.\ \cite{meylan}), which means that the cluster members are well
separated from the field stars. In an effort spanning twelve years,
Mayor, Meylan and co-workers obtained {\tt CORAVEL} radial velocities
for 471 cluster members extending to about 20$'$, with mean errors of
better than 1 \kms\ (see Mayor et al.\ \cite{mm_data} and
\S\ref{rv_S:external} below). Their data showed that $\omega$~Cen
rotates, with the mean rotation reaching a peak of about 8 \kms\ at
$8'$ from the cluster centre (e.g., Merritt, Meylan \& Mayor
\cite{merritt}).  Suntzeff \& Kraft (\cite{sk96}) used {\tt ARGUS} to
obtain radial velocities of somewhat more modest accuracy for 360
members on the giant branch between $3'$ and $23'$.  Given the very
large number of stars for which proper motions are now available, and
given the clean member selection possible with radial velocities, we
embarked on a radial velocity study with {\tt ARGUS} to extend the
published datasets significantly.\footnote{Radial velocities were
obtained for nearly 5000 stars in the central three arcmin of
$\omega$~Cen with the Rutgers imaging Fabry-Perot spectrometer (Xie,
Gebhardt, et al., in prep.)} Here we present the resulting radial
velocities.\looseness=-2

In \S\ref{rv_S:obs} we briefly describe the {\tt{ARGUS}} instrument,
our observations and data reduction, a comparison with earlier work,
and the results of an extensive error analysis.  We briefly discuss
the results in \S\ref{rv_S:results}: membership determination,
rotation and velocity dispersion. The conclusions follow in
\S\ref{rv_S:concl}. In the companion Paper III (van de Ven et al.\
\cite{vbv+05}), we use our radial velocities to correct the Paper I
proper motions for remaining overall (solid body) rotation, correct
both the radial velocities and the proper motions for the perspective
rotation caused by the space motion of $\omega$ Cen, and compare the
resulting internal motions with anisotropic axisymmetric dynamical
models, to derive an accurate dynamical distance for the
cluster.\looseness=-2

\section{Observations}
\label{rv_S:obs}

The observations were made with the CTIO 4m Blanco telescope during 16
nights, on February 11--14 1992, March 1--6 1993, and February 26 --
March 3 1994.  The seeing was better than 1.5 arcsec FWHM for most of
the data.  We used the fiber-fed, multiple-object echelle spectometer
{\tt{ARGUS}} to obtain a total of 2756 radial velocity measurements
for 1966 individual stars (see also Appendix A).

\subsection{The Instrument}

{\tt ARGUS} resided at the prime focus of the Blanco telescope. It
consisted of 24 computer-controlled arms located around the
telescope's f/2.66 prime focus field.  Use of the red doublet
corrector produced a flat field of 46$'$ in diameter at a scale of
18.6 arcsec/mm.  The fibers were 100~$\mu$m or 1\farcs86 in diameter
and carried light from the prime focus cage to a spectrograph located
in a thermally and mechanically isolated room. The movement of the
fibers was in units of `steps' where a single step is 10~$\mu$m or
0\farcs2. {\tt{ARGUS}} had the ability to rapidly change the
configuration of the fibers; and it had a high dispersion ($6.7
{\hbox{\AA}}\;\hbox{mm}^{-1}$ at Mg{\it b}) which made accurate
radial velocity measurements possible. For more details on {\tt
ARGUS}, see {\sl{www.ctio.noao.edu/spectrographs/argus/argus.html}}
and Ingerson (\cite{ingerson}), Lutz et al.\ (\cite{lutz}).

The {\tt ARGUS} echelle mode employed a 31.6 l/mm echelle grating. We
used an order-separating filter to isolate a single echelle order
centred on the Mg{\it b} triplet near $5175$\AA.  This wavelength
region also contains numerous other sharp lines and is an ideal and
much used region for accurate radial velocity measurements of
late-type stars.\looseness=-2

In 1992 and 1994 we used a Reticon CCD in the blue Air Schmidt camera.
It had $1240\times400$ pixels of 27$\mu$m size, and a read noise of
2.94~$e$. The wavelength range covered was 5083--5281 \AA\ in 1992,
and 5081--5274 \AA\ in 1994, with a spectral resolution of 0.3 \AA\
(4.2 \kms/pixel).  Due to technical difficulties with this camera in
1993, we were forced to use a GEC CCD in the red Air Schmidt
camera. The GEC CCD had $576\times425$ pixels of size 22 $\mu$m, and a
read noise of 5.18~$e$. The wavelength range covered was 5133--5218
\AA, and the resolution was 0.6 \AA\ (8.9 \kms/pixel).

In general the instrument worked well, although we were troubled by
the poor pointing accuracy of the fibers during all three years.  This
required every fiber to be manually centred on a star.  During 1994 we
were observing at full moon, which limited how faint we could
observe. Exposure times ranged from 600 to 1800 seconds.

Our observing strategy was to do as many stars as possible in one
region of the cluster before changing the centre of the field.  In
each region we selected two bright stars (not necessarily cluster
members) as local standards, and locked two fibers on these stars for
all observations in this field.  This allowed us to monitor the drift
of the system without the expense of doing comparison arcs after every
exposure, which were done instead at the start and end of every field
(before and after any telescope motion). Comparison arcs were also
done either before or after any standard star or twilight sky
measurement. In 1992 we used HD~31871 ($v_{\rm helio}=62.3$ \kms) and
HD~43880 ($v_{\rm helio}=46.9$ \kms) as radial velocity standards. In
1993 and 1994 we used HD~120223 ($v_{\rm helio}=-26.3$ \kms) and
HD~176047 ($v_{\rm helio}=-42.8$ \kms).

\subsection{Star Selection}

We originally selected the stars to be observed from the {\em{initial
preliminary}} proper motion catalogue (see Paper~I, \S4), mainly at a
distance $r \geq 3.5'$ from the cluster centre.  All photographic
plates that were used for that study were centred on the same position
and the cluster centre is not in the centre of the plates. This
enabled us to study stars that are relatively far ($>30'$) from the
cluster centre at least in one direction.  The {\em final} proper
motion catalogue contains cluster giants, horizontal branch stars and
field stars down to a photographic magnitude $B=16$ in the cluster
centre and 16.5 in the outer regions. Table 5 of Paper I lists their
properties, including $B$ magnitude and $B-V$ colour (when
available). Because we selected objects from the initial catalogue
which was somewhat larger than the final catalogue, we measured radial
velocities of some stars that are not present in Table 5 of
Paper~I. We discuss these in Appendix A.\looseness=-2

In Paper I, each star was assigned a class on a scale from 0 to 4,
based on visual inspection of the stellar image on the photographic
plates, and ranging from no disturbance by a neighbouring star (0) to
badly disturbed (4). Over 83\% of our stars are of class 0, and the
remainder is divided nearly equally between classes 1 and 2 (mildly
disturbed).

In selecting stars for radial velocity measurement, we wanted to be
careful not to introduce any kinematic bias. Thus we avoided selecting
stars on the basis of any proper motion value, in order to avoid
truncating the observed radial velocity distribution function.

We also elected not to select on the basis of location in the
colour-magnitude diagram, other than that imposed by the lack of
spectral lines in blue stars. In what follows, we restrict ourselves
to stars with $B-V\geq 0.4$. We specifically did not exclude stars
from the sample which did not fall in the giant and sub-giant region
of the colour-magnitude diagram, in order to have as complete a sample
as possible. The price of not imposing such selection criteria is that
we ended up observing a considerable number of field stars. In
Appendix A we return to the stars with $B-V<0.4$ and those without
colour information.\looseness=-2

In 1992, instrument problems forced us to work in the `inner' region
($3\farcm5 < r < 8\farcm0$) only, and we obtained 649 observations of
564 stars. In 1993 and 1994 we worked in both the inner and outer
regions (from $3\farcm5$ to $38'$), and obtained 1673+944 observations
of 1256+707 stars.

\subsection{Data Reduction}
\label{rv_S:data_reduction}

After trimming the raw images and correcting for overscan and fixed
pattern bias, we divided the spectra of 1993 and 1994 by a 'milk' flat
field to remove pixel-to-pixel variations (see {\tt ARGUS} web page).
The different spectra on each frame were identified, traced and
extracted using the IRAF {\tt APSUM} task. For the 1992 data, we ran
{\tt APSUM} before we applied the flat field corrections, because we
used a quartz lamp flat field. After continuum subtraction, we
rebinned the spectra onto a log wavelength scale.

We used the IRAF {\tt RV} package (Tody \cite{tody}) to
cross-correlate the spectra against high S/N template spectra of two
radial velocity standard stars of similar spectral type (Tonry \&
Davis \cite{tonry}).  The displacement of the peak of the correlation
function gives the velocity of the star relative to the template. We
filtered out obvious bad velocities by noting the value of the cross
correlation coefficient below which the scatter in repeat observations
increased significantly. As reported by C\^ot\'e et al.\ (1994), the
correlation task {\tt FXCOR} in IRAF tends to overestimate the
uncertainties in the derived radial velocities by as much as a factor
of two.\looseness=-2

\begin{table}
\caption[]{Mean radial velocities and standard errors for stars with
           more than 10 repeat measurements. The table gives: year of
           observation; Leiden identification number from Paper I; ROA
           number from Woolley (1966); photographic $B$-magnitude of
           Paper I; number of measurements $n$; mean velocity $\langle
           v \rangle$; standard deviation $\sigma_{\langle v \rangle}$
           of the mean velocity.  }
\label{rv_T:repeat}
\begin{center}
\begin{tabular}{cccccrc}\hline\hline\noalign{\smallskip}
Year   & star   & $\!\!$ROA &$B$ & $n$
                            &$\langle v \rangle\ \ \null$
                            &$\!\!\sigma_{\langle v \rangle}$ \\
       &   nr      & nr & mag  &  & {\scriptsize{\kms}} & {\scriptsize{\kms}}
                                                \\ \noalign{\smallskip}\hline
1992   &  46024 & 40 & 12.77 & 47 &    215.95& 1.20 \\
       &  60087 & 20 & 11.08 & 38 &  $-$23.04& 0.26 \\
1993   &  62015 & 36 & 12.66 & 22 &  $-$30.30& 1.08  \\
       &  93011 & -  & 12.47 & 23 &  $-$26.25& 2.61  \\
1993/4 &  11014 & 242& 12.54 & 53 &  9.45    & 1.59  \\
1994   &  27009 & 409& 13.22 & 17 &  46.42   & 0.81  \\
       &  32138 & 48 & 12.98 & 20 &  222.08  & 0.70  \\
       &  48049 & 76 & 13.04 & 20 &  219.16  & 1.28  \\
       &  65014 & 413& 13.80 & 17 &  $-$44.90& 1.60  \\
\noalign{\smallskip}\hline
\end{tabular}
\end{center}
\end{table}

\subsection{Repeat Measurements}
\label{rv_S:error}

A number of factors contribute to the uncertainty of the velocities:
the errors in the individual measurements, the error in the
fiber-to-fiber velocity zero point, and the zero point of the velocity
system with respect to the standard stars. In addition, the radial
velocity of some stars will vary in time due to orbital motion in a
binary.

We carried out simple Monte Carlo simulations using 18 bright known
cluster members. These stars were measured in one frame and, before
processing, we copied this frame 24 times and added random noise to
each frame using the IRAF {\tt MKNOISE} task, simulating the same
conditions (read noise, gain etc.) as the real data.  In this way, we
simulated 24 measurements of 18 bright stars. We found a standard
deviation $\sigma_{\rm sim} \lta 1 $ \kms.

Stars with 10 or more measurements are listed in
Table~\ref{rv_T:repeat}.  Their standard errors range from less than
0.3 to 2.6 \kms. Column 1 indicates the year in which the observations
were taken; columns 2 and 3 contains the star identification numbers
in our catalogue and the corresponding ROA number from the survey of
Woolley (\cite{woolley}); column 4 the star's photographic $B$
magnitude (Table 5 of Paper I); column 5 gives the number of repeat
measurements; columns 6 and 7 the mean radial velocities $\langle
v\rangle$ and their standard errors $\sigma_{\rm \langle v \rangle}$.
The calculated standard deviation from repeat measurements agrees
reasonably well with the standard deviation found from the
simulations, but it is on average 0.8 times the formal (standard)
error provided by {\tt FXCOR}. We return to this result in
\S\ref{rv_S:external}.\looseness=-2

\begin{figure}[t]
\begin{center}
\includegraphics[height=19.6truecm]{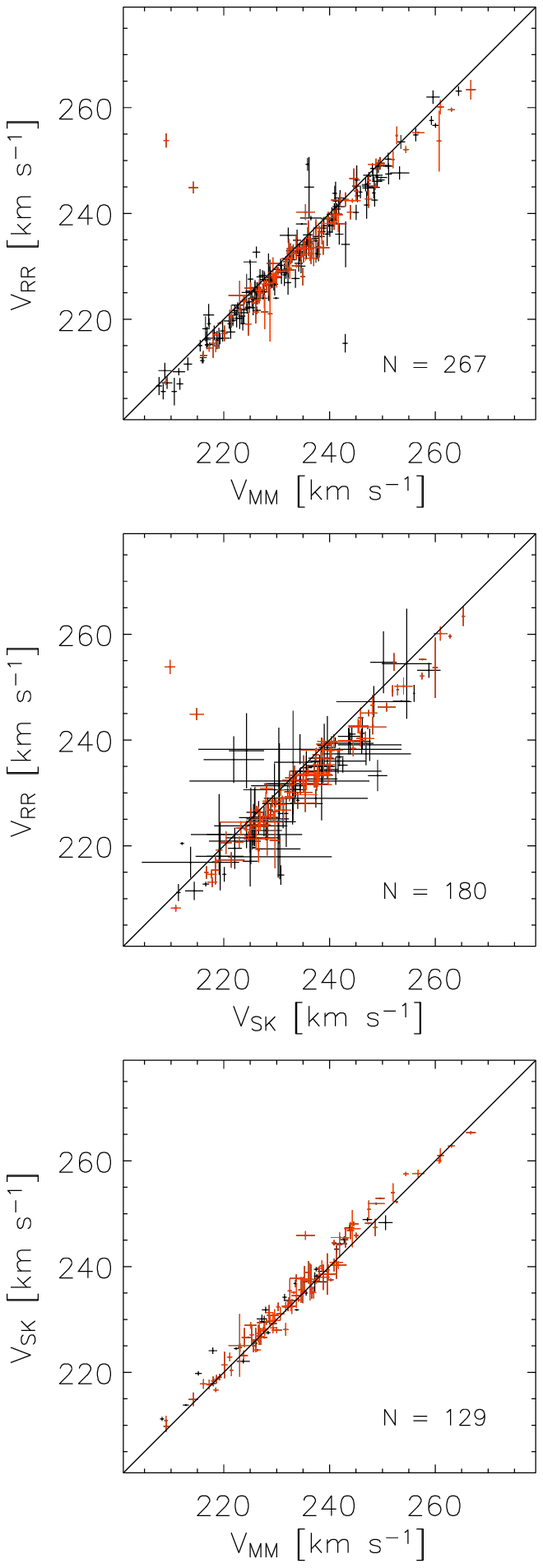}
\caption{Comparison of our radial velocity measurements ($rr$) with
those of Mayor et al.\ (1997, $mm$) and of Suntzeff \& Kraft (1996,
$sk$). a) $rr$ versus $mm$, b) $rr$ versus $sk$, and c) $sk$ versus
$mm$, for stars in common. The red symbols identify the subset of 100
stars in common to all three studies. }
\label{rv_F:diff}
\end{center}
\vskip -0.6truecm
\end{figure}

Stars with a large number of repeat measurements were always measured
with the same fibers. We did similar Monte Carlo simulations to track
down uncertainties and differences between the individual fibers using
twilight spectra. Those differences are also well within the standard
errors.  The velocities of stars that were measured two or three
times, but with different fibers and at different nights and even in
different observing runs in general also agreed well within the
standard errors.

There are ten stars which we observed in all three years. 72 stars
overlap 1992--1993; 46 in 1993--1994; and 108 in 1992--1994. These
measurements are in good agreement (the differences are smaller than
the estimated uncertainties) except for a few probable
misidentifications or binaries.  Some misidentifications are probably
due to pointing problems of some fibers and these measurements were
not included. Some of the stars with a high number of repeat
measurements ($n\ge10$) are bright field stars.

The observations in 1993 were carried out with a setup that differed
from that in 1992 and 1994 (see \S 2.1). The repeat measurements show
that the best radial velocities from 1993 have errors as small as
those from 1992 and 1994. However, the 1993 error distribution has a
slightly more extended tail to larger values. Because many stars had
repeat measurements in two years, we decided to treat the entire data
set as one sample.

A physical source of velocity differences in the repeat measurements
is the presence of binaries.  Double stars or multiple stars will show
a larger dispersion due to their orbital motion. From all probable
cluster members (see \S\ref{rv_S:mem}) we have 303 stars measured more
than once. Only about $4\%$ of these have repeat measurements that
differ significantly from the uncertainties indicated by the IRAF
tasks $(\sigma_{\rm{rep}}>1.5 \times\sigma_{\rm{iraf}})$. This is an
indication that there is only a small fraction of binaries with
short-periods in the cluster. Assuming that $\omega$~Cen has a period
distribution similar to the one observed for nearby G dwarfs, Mayor et
al.\ (\cite{mayor96}) estimated the global binary frequency for binary
systems with periods less than $10^4$ days in $\omega$~Cen to be as
low as 3--4\%.

\subsection{Comparison with Other Studies}
\label{rv_S:compare}

Mayor et al.\ (\cite{mm_data}) published radial velocities of 471
stars between $10''$ and $1342''$ from the cluster centre, obtained
with {\tt CORAVEL}. They found a mean radial velocity of the cluster
of $\langle v_{mm}\rangle = 232.8 \pm 0.7$ \kms.  We have compared
individual velocities of our sample (`$rr$') with theirs (`$mm$'). We
have 267 stars with ROA numbers in common. Figure~\ref{rv_F:diff}a
shows the velocities plotted against each other with their
uncertainties.  The velocities generally agree well, but there is a
small systematic offset, which we ascribe to a zero-point error in our
data. There are some outliers, notably two stars at $v_{mm}\approx
210$ \kms (LID 44065, 60065) which we assume are misidentifications
(see below), and LID 35090 and 78035 where our measurements differ by
nearly 220 \kms\ from the values reported by Mayor et al. We removed
outliers by discarding all stars for which the measured velocities
differ by more than four times the combined one-sigma error
(`four-sigma clipping'). This leaves 250 stars. The weighted mean
velocity offset between the two studies is $\langle{v}_{rr} -
{v}_{mm}\rangle = -1.44\pm0.09$ \kms.

Suntzeff \& Kraft (\cite{sk96}) observed 199 members of $\omega$~Cen
with $M_V\sim 1.25$ on the lower giant branch at radial distances
between $8'$ and $23'$ (`faint sample'), and 144 members at $M_V\sim
-1.3$ at radial distances between $3'$ and $22'$, to which they added
another 17 observed by Seitzer (`bright sample').  They measured the
velocities with {\tt ARGUS}, but in the wavelength range $8200 -
8800$~\AA\ using the Ca {\tt II} triplet.  We have 180 stars in common
with their total sample of 360 stars.  Figure~\ref{rv_F:diff}b shows
the $rr$ velocities versus the $sk$ measurements. There is one star at
about $v_{rr}\approx 220$ \kms\ which has been measured very precisely
by both groups, but the measured velocities differ significantly (8
\kms).  This large difference could be due to a chance combination of
large errors, a binary, or is a misidentification. The two outliers at
$v_{sk}\approx 210$ \kms\ are the same as those noted in
Figure~\ref{rv_F:diff}a, confirming that they are misidentifications
on our side. After four-sigma clipping we are left with 172 stars in
common, which have a mean velocity offset of
$\langle{v}_{rr}-{v}_{sk}\rangle = -2.02\pm0.15$ \kms.

Figure~\ref{rv_F:diff}c shows the comparison of the $sk$ and $mm$
velocities. These data set have 129 stars in common. Four-sigma
clipping reduces this to 117, with mean offset
$\langle{v}_{sk}-{v}_{mm}\rangle=0.41\pm0.08$ \kms.

Finally, we have two stars in common with Tyson \& Rich
(\cite{tyson}), namely ROA 55 ($v_{rr}=221.7\pm0.9$ \kms,
$v_{tr}=220.4\pm0.2$ \kms) and ROA 70 ($v_{rr}=227.7\pm2.7$ \kms,
$v_{tr}=213.9\pm4.4$ \kms). Both stars were observed multiple times by
Mayor et al.\ (\cite{mm_data}) with an accuracy of better than 1 \kms,
and are without a doubt variable: 39 measurements of ROA 55 give
$v_{mm}=226.3\pm4.0$ \kms\ and 23 measurements of ROA 70 give
$v_{mm}=229.9\pm4.3$ \kms.

\begin{table}[t]
\caption[]{Comparison of three independent studies. Mean and
dispersions of the pairwise differences in measured radial velocities
for the 93 stars in common between the $rr$, $mm$, and $sk$ samples,
after four-sigma clipping to remove outliers. }
\label{rv_T:diff}
\begin{center}
\begin{tabular}{cccc}
\hline\hline\noalign{\smallskip}
      &  $rr-mm$ & $rr-sk$ & $sk-mm$ \\
\noalign{\smallskip}\hline\noalign{\smallskip}
$\langle v\rangle $ (\kms) & $-1.50\pm 0.12$ & $-2.09\pm 0.17$& $0.37\pm0.10$\\
$\sigma$ (\kms) & $\phantom{-}1.26\pm0.09$
                & $\phantom{-}1.68\pm0.12$ & $1.37\pm0.07$\\
\noalign{\smallskip}\hline
\end{tabular}
\end{center}
\end{table}

\begin{table}[t]
\caption[]{Median and external measurement errors for the three
studies ($rr$, $sk$ and $mm$) described in the main text. The second
and third columns list the median standard error $\sigma_{\rm median}$
and the external error $\sigma_{\rm external}$ for the 93 stars in
common to the three studies. The fourth column lists the median error
for the total sample, and the fifth column the inferred external error
(see main text). The units are \kms.}
\label{rv_T:external}
\begin{center}
\begin{tabular}{ccccc}
\hline\hline\noalign{\smallskip}
Study  & \multicolumn{2}{c}{93 overlap stars}  &
          \multicolumn{2}{c}{total sample } \\
     &$\sigma_{\rm median}$ &$\sigma_{\rm external}$
                    &$\sigma_{\rm median}$ &$\sigma_{\rm external}$ \\
\noalign{\smallskip}\hline\noalign{\smallskip}
$rr$ &  1.43 & $1.13\pm0.08$   & 2.65  &$2.09\pm0.15$ \\
$sk$ &  0.90 & $1.25\pm0.08$   & 1.60  &$2.22\pm0.14$ \\
$mm$ &  0.56 & $0.56\pm0.08$   & 0.61  &$0.61\pm0.09$ \\
\noalign{\smallskip}\hline
\end{tabular}
\end{center}
\end{table}

\begin{figure}[t]
\begin{center}
\includegraphics[width=0.45\textwidth,height=17.0truecm]
                 {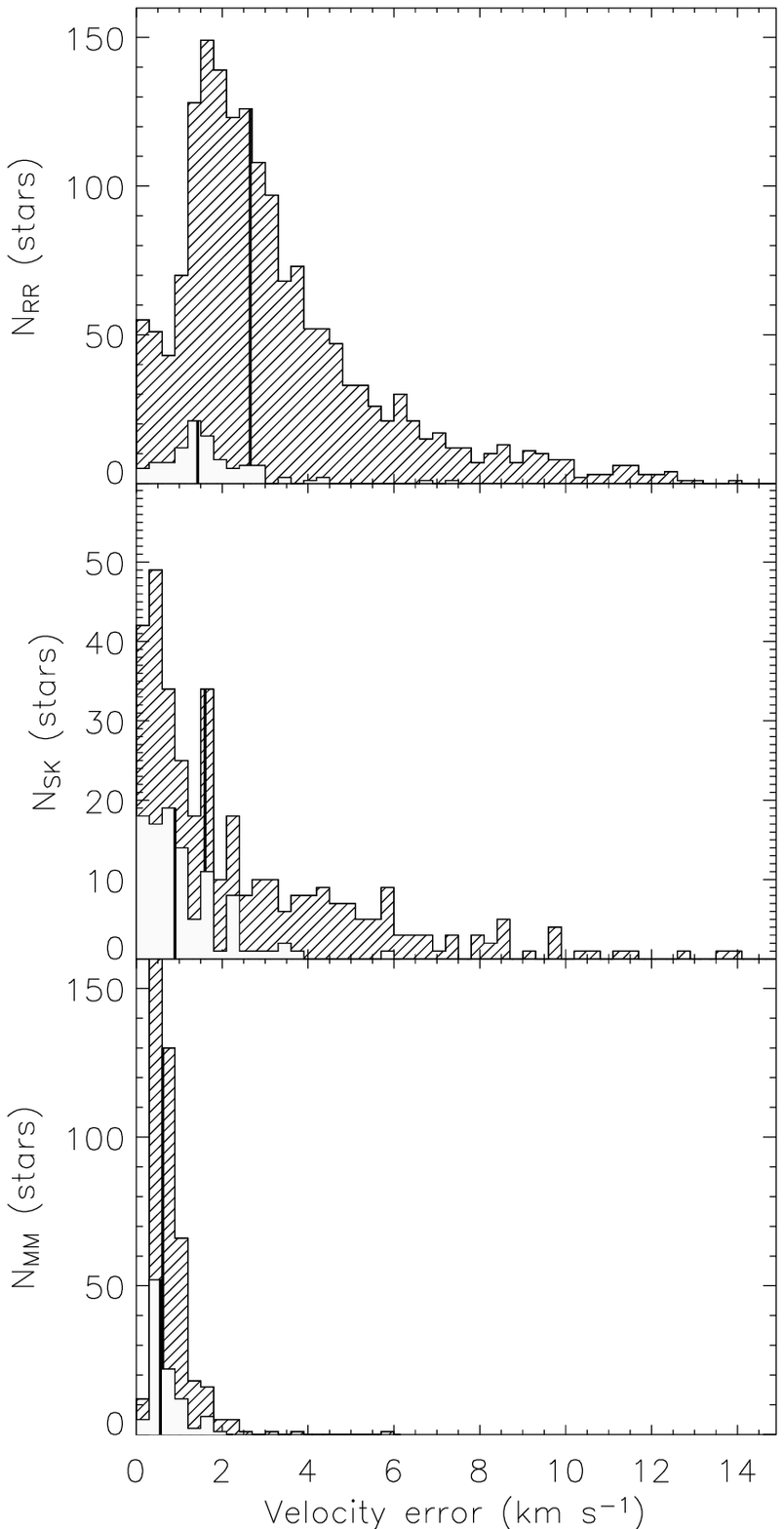}
\caption{Histograms of errors in the radial velocities reported for
         the $rr$, $sk$ and $mm$ samples (top, middle and bottom
         panel). The light-shaded areas are the similar histograms for
         the 93 stars in common between the three studies. The solid
         vertical lines indicate the median value of each of the
         distributions.}
\label{error_hist}
\end{center}
\vskip -0.6truecm
\end{figure}

\subsection{External Error Estimate}
\label{rv_S:external}

There are 100 stars in common between all three studies ($rr$, $mm$
and $sk$). Removing outliers by means of four-sigma clipping leaves
93, which cover the full range of velocities seen in $\omega$ Cen
(Fig.~\ref{rv_F:diff}). Table~\ref{rv_T:diff} lists the mean and the
dispersions of the pairwise differences, as well as the dispersion in
these values, for these 93 stars, calculated with the expressions
summarized in Appendix B. As expected, the differences are fully
consistent with the systematic offsets between the three studies
derived in \S\ref{rv_S:compare} for the larger samples of pairwise
overlaps.

\begin{figure}[t]
\begin{center}
\includegraphics[width=0.45\textwidth]
                 {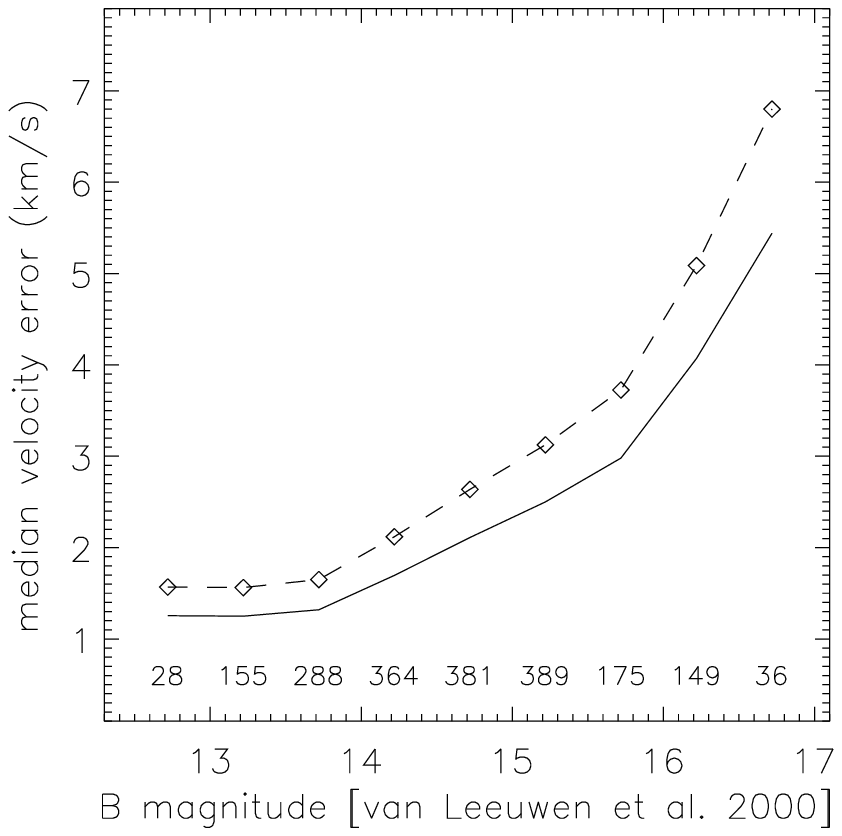}
\caption{Median error in our radial velocity measurements as a
         function of photographic $B$-magnitude, in 0.5 magnitude
         bins. The numbers below the diamonds indicate the number of
         measurements in each bin. The dashed line connects the median
         error derived from the standard errors provided by {\tt
         FXCOR}. Our external error analysis demonstrates that these
         standard errors are overestimated, and need to be multiplied
         by a factor 0.8. The solid line shows the corrected median
         errors, which correspond to the individual errors reported in
         Table~\ref{rv_T:summary}.\looseness=-2 }
\label{median_error}
\end{center}
\vskip -0.3truecm
\end{figure}

The dispersions listed in Table~\ref{rv_T:diff} allow us to estimate
the true external sigmas for each of the three studies. Since
$\sigma_{v_1-v_2}^2 = \sigma_{v_1}^2 + \sigma_{v_2}^2$ etc., we can
solve for the true $\sigma_{v_i}$ ($i=1, 2, 3$) given the three
$\sigma_{v_i-v_j}^2$'s. The results are listed in the third column of
Table~\ref{rv_T:external}, and can be compared with the median of the
reported individual errors for these 93 stars in each of the three
studies. This reveals that the {\tt CORAVEL} errors are estimated
accurately, but that there are discrepancies for the {\tt ARGUS}
velocities. The errors reported by $sk$ appear to be 0.7 times the the
external error, while the external errors in our own study appear to
be 0.8 times the formal errors provided by {\tt FXCOR}. This is not
unexpected, as {\tt FXCOR} is known to overestimate the errors (see
\S\ref{rv_S:data_reduction}).\looseness=-2

Figure~\ref{error_hist} shows the histograms of the reported errors in
the radial velocity measurements for all three studies, not only for
the total samples, but also for the 93 stars in common to all three
studies. All error distributions are skewed, with particularly
significant tails towards large errors for the {\tt ARGUS} velocities.
The median errors $\sigma_{\rm median}$ listed in
Table~\ref{rv_T:external} are indicated by vertical solid lines. The
93 overlap stars are all brighter than $B$-magnitude of 14, and as a
result their errors are systematically smaller. This suggests that the
average true external errors for the complete samples are larger than
estimated from just the 93 overlap stars. We estimate these errors by
simply multiplying the median error with the same correction factor as
found for the 93 stars. The results are listed in the fifth column of
Table~\ref{rv_T:external}, and show that our {\tt ARGUS} data for the
Mg{\it b} wavelength region are of the same accuracy as the {\tt
ARGUS} measurements by Suntzeff \& Kraft (\cite{sk96}) around the
Ca~{\tt II} triplet, and are about a factor two less accurate than the
{\tt CORAVEL} data of Mayor et al.\ (\cite{mm_data}).\looseness=-2

The fibers we used for our observations are relatively large (\S 2.1),
and might sometimes contain a contribution of a neighbouring star, or
of unresolved background light. This could cause the derived radial
velocities to be biased towards the mean cluster velocity, which would
also bias the derived velocity dispersion. This systematic effect is
expected to be stronger in the crowded center, and for fainter stars.

Our sample of 1966 stars contains 1634 stars without any neighbour
disturbing the stellar image (class 0 stars of Paper I).  None of the
remaining 182$+$150 class 1 and 2 stars, the images of which are at
most mildly disturbed, overlap with the $sk$ or $mm$ samples, so no
external error estimate is possible for this subset of our sample.  We
have computed the mean velocity and intrinsic velocity dispersion for
each of these two classes, and find no difference with the similar
results for the class 0 subsample. Furthermore, plots of the
differences $v_{rr}-v_{mm}$ and $v_{rr}-v_{sk}$ versus magnitude and
versus distance to the centre show no evidence for any systematic
trends. This is particularly relevant for the comparison with the $sk$
sample, which consists of a `bright' and a `faint' subset, together
covering nearly the entire magnitude range of our sample. Finally, we
also checked whether the intrinsic dispersion for the entire set
depends on magnitude in different radial bins.  We experimented with
the size and location of the radial and magnitude bins, but other than
the expected decline of velocity dispersion with radius (see Paper
III), found no evidence for a dependence on magnitude. We conclude
that effects of crowding by neighbours and of unresolved background
light can be ignored safely, and that our external error estimate is
valid for the entire sample.\looseness=-2

\begin{figure}
\begin{center}
\includegraphics[width=0.47\textwidth,height=!]{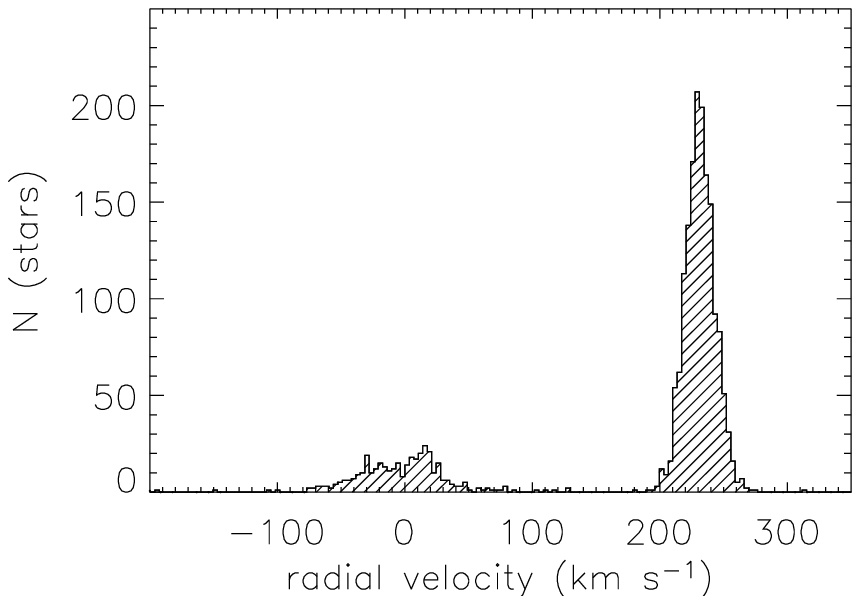}
\caption{Histogram of the 1966 measured radial velocities in the
         direction of $\omega$ Centauri. The distribution centred
         around radial velocity zero corresponds to the field stars,
         and the strongly peaked distribution centered near 232 \kms\
         is the cluster.}
\label{rv_F:hist}
\end{center}
\end{figure}

Figure~\ref{median_error} shows the median standard error of our
measurements, as provided by {\tt FXCOR}, plotted versus photographic
$B$-magnitude (Paper I), in bins of 0.5 magnitude. The number of stars
per bin is indicated also. The measurement for LID 60087 with
$B=11.08$ (see Table~\ref{rv_T:repeat}) is not included. Our external
estimate suggests that the real errors are smaller by a factor 0.8, so
we also show the same curve multiplied by this factor. The resulting
median errors are below 1.5 \kms\ for stars brighter than $B=14$,
increase to 2.5 \kms\ at $B=16$, and continue to increase for the
modest number of fainter stars.\looseness=-2

Table~\ref{rv_T:summary} presents our results.  Column 1 gives the
Leiden identification number of the star (LID; Paper I). Column 2
gives the mean heliocentric radial velocity $\langle v \rangle$ and
Column 3 lists our best estimate of the standard error
$\sigma_{\langle v \rangle}$ (0.8 times the standard error provided by
{\tt FXCOR}).  Column 4 gives the number of individual {\tt ARGUS}
measurements for each star. The associated coordinates $\alpha$ and
$\delta$, and the $B$ magnitude and $B-V$ colour, can be found in
Table 5 of Paper I. As discussed in \S~2.5, the reported measurements
for LID 35090, 44065, 60065, and 78035 are suspect.

\begin{table}
\caption[]{Summary of results, extract only. LID: Leiden
identification number from Paper I; mean measured heliocentric radial
velocity $\langle v \rangle$ and the standard error $\sigma_{\langle v
\rangle}$, in \kms; $n$: number of individual measurements (see text
for details). }
\label{rv_T:summary}
\begin{center}
\begin{tabular}{cccccc}\hline\hline\noalign{\smallskip}
LID  &$\langle v \rangle$ & $\sigma_{\langle v \rangle}$  &$n$ \\
     &  \kms  & \kms &  &       \\
\noalign{\smallskip}\hline\noalign{\smallskip}
 00009 & 208.0  & 1.2 & 1 \\
 00012 & 242.2  & 3.3 & 1 \\
 00014 & 219.6  & 6.0 & 1 \\
 01010 &  26.6  & 5.7 & 1 \\
 01015 & 235.2  & 2.0 & 1 \\
\noalign{\smallskip}\hline
\end{tabular}
\end{center}
\end{table}

\section{Results}
\label{rv_S:results}

We briefly discuss membership and distribution over the cluster for
our sample of stars, present a colour-magnitude diagram, and describe
the mean velocity and velocity dispersion field. We use these as input
for the determination of the dynamical distance to $\omega$~Cen in
Paper III.\looseness=-2

\subsection{Membership Determination}
\label{rv_S:mem}

Figure~\ref{rv_F:hist} shows a histogram of all 1966 mean radial
velocities. Because of the large radial velocity of the cluster
($232.8\pm0.7$ \kms, Mayor et al.\ \cite{mm_data}) there is a clear
distinction between cluster members and field stars. Our sample
contains 1589 stars with a velocity between 160 and 300 \kms. Paper I
reports proper motion measurements for all of these, and notes that
36 have a low probability of being a member based on the proper
motions alone. The separation between cluster and field is
sufficiently clean in radial velocity to consider all 1589 stars
secure members.

\begin{figure}
\begin{center}
\includegraphics[width=0.47\textwidth,height=!]{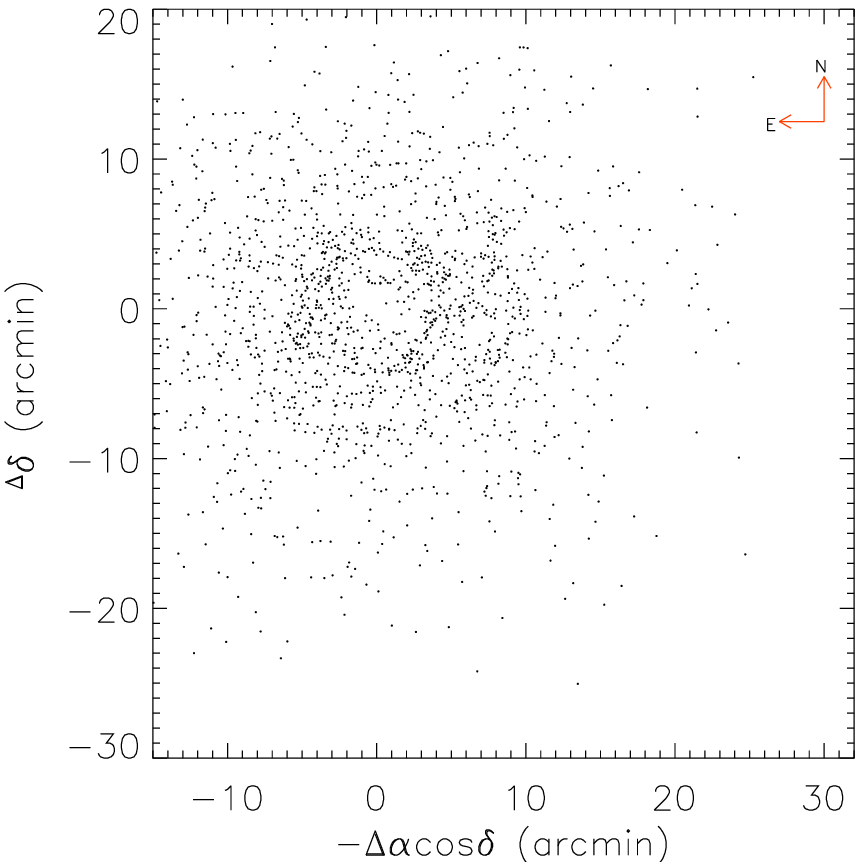}
\caption{Positions of 1589 possible radial velocity members which have
         proper motions (Paper I), relative to the centre of $\omega$
         Centauri. }
\label{rv_F:xy}
\end{center}
\end{figure}

Figure~\ref{rv_F:xy} shows the positions for the 1589 radial velocity
members with proper motion measurements. Since the cluster centre does
not coincide with the centre of the plate material from which the
proper motions were derived, we have full azimuthal coverage only up
to $15'$ from the cluster centre. Stars with a radial velocity
measurement are fairly evenly spread over the cluster, except for a
stronger concentration of measurements in a ring at about $4'$, which
is due to observational constraints.  The furthest member based on
radial velocity measurements is star 95002 (no ROA number available)
with a radial velocity of $209.1$ \kms, at a distance of $37\farcm$7
(0.8 $r_t$) from the cluster centre.

\begin{figure}
\begin{center}
\includegraphics[width=0.47\textwidth,height=!]{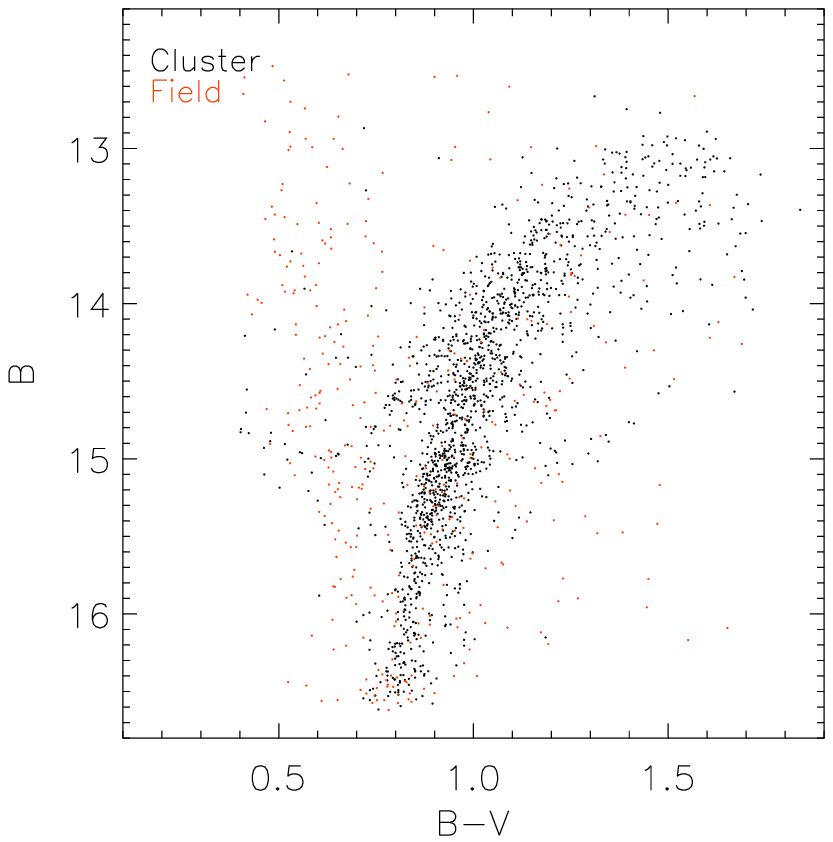}
\caption{Colour-magnitude diagram for the subset of 1966 stars for
         which Paper I reports a $B$-magnitude and $B-V$ colour. Of
         these, 1589 have a heliocentric radial velocity in the range
         between 160 and 300 \kms, and are considered secure members
         of $\omega$ Cen. These are indicated by the black dots. The
         remaining 504 stars (red dots) are field stars. }
\label{rv_F:cmd}
\end{center}
\end{figure}

\subsection{Colour-Magnitude Diagram}
\label{rv_S:cmd}

Fig.~\ref{rv_F:cmd} shows the colour-magnitude diagram for all stars
from our sample of 1966 for which Paper I gives a $B$-magnitude and
$B-V$ colour. The colour-excess $E(B-V)$ towards $\omega$ Cen has been
established as 0.11 mag (Lub \cite{lub02}). Most cluster members are
found on the giant branch, which is quite broad, and extends redwards,
most likely as the result of the presence of multiple stellar
populations in the cluster (e.g., Norris \& Da Costa \cite{nordca};
Pancino \cite{pan02}). The `anomalous giant branch' (Lee et al.\
\cite{ljs+99}; Pancino et al.\ \cite{pfb+00}; Ferraro et al.\
\cite{fsp+04}) corresponds to the detached giant branch passing
through $B = 15$ and $B-V = 1.2$, containing about two dozen
stars. Their mean radial velocity and velocity dispersion are
consistent with the values for the entire cluster. The few bright
members with $B<13$ bluewards of the giant branch, reminiscent of
Fehrenbach's star HD116745 (LID 16018; see Fehrenbach \& Duflot 1962),
are possibly post-AGB stars. The non-members are distributed more
homogeneously over the diagram, as expected. The near-vertical band
with $B-V\sim 0.55$ consists of main-sequence stars over a significant
range of distances.

\begin{figure}[t]
\begin{center}
\includegraphics[width=0.50\textwidth]{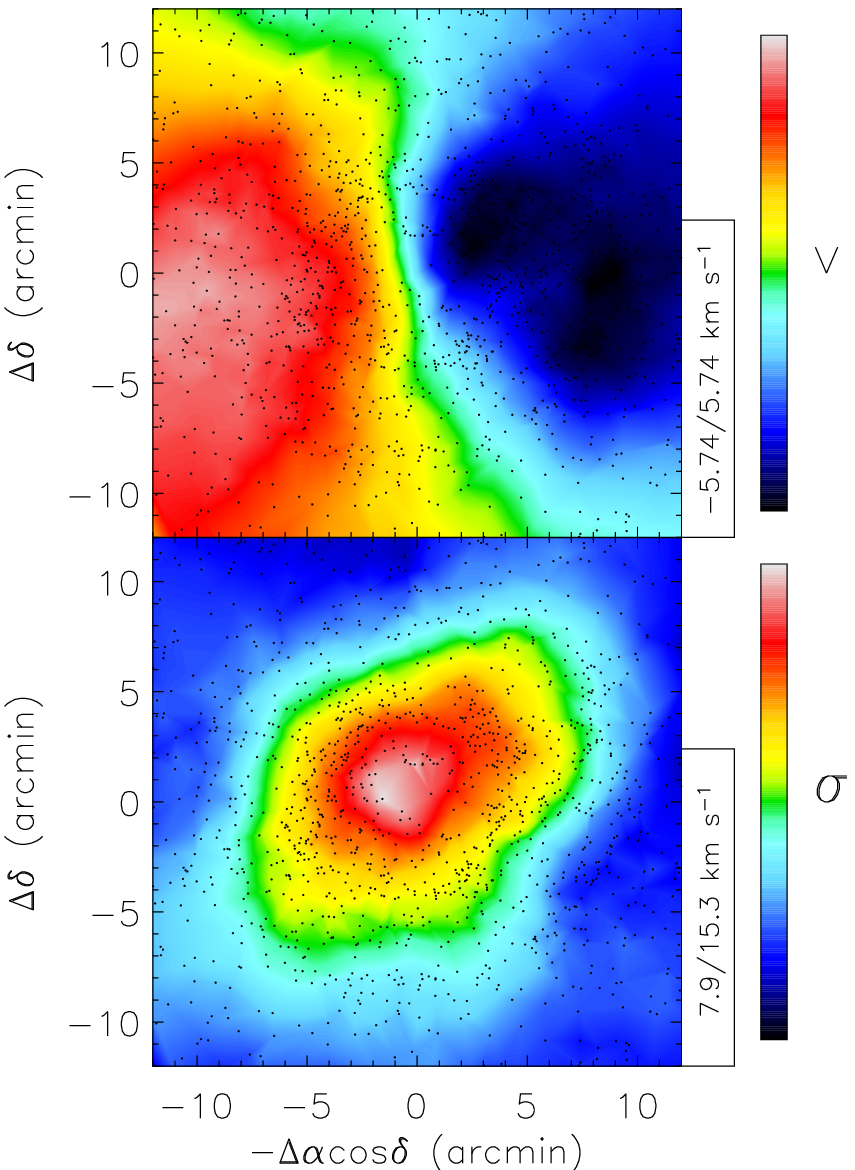}
\caption{Kinematics of $\omega$~Cen, based on adaptive smoothing of
the individual measurements, as described in the text.  a) mean radial
velocity field after subtraction of the systemic velocity, b) radial
velocity dispersion.  The dots indicate the positions of the
individual measurements. The observed velocities have not been
corrected for the perspective rotation caused by the space motion of
the cluster, and these maps should therefore be interpreted with
caution (Paper III).\looseness=-2}
\label{rv_F:v_field}
\end{center}
\end{figure}

\subsection{Rotation and Velocity Dispersion}
\label{rv_S:rotation}

An indication of internal rotation of $\omega$~Cen based on proper
motion measurements was found by Dickens \& Woolley (\cite{dw67}).
They saw evidence for solid body rotation in the centre of the cluster
and showed that the apparent ellipticity of $\omega$~Cen $(\varepsilon
= 1-b/a)$ increases from approximately zero in the centre to a maximum
of 0.25 at $10'$, and decreases slowly and regularly to almost zero at
the tidal radius of the cluster. Dickens \& Woolley (\cite{dw67})
suggested that the rotation axis should be approximately North-South
in the plane of the sky.  These conclusions were confirmed by Geyer,
Hopp \& Nelles (\cite{geyer2}).  Rotation was observed as well by
Harding (\cite{harding}) based on radial velocities of 13 stars in the
cluster, and was later confirmed by the {\tt CORAVEL} study (Meylan et
al.\ \cite{meylan}; Merritt et al.\ \cite{merritt}).\looseness=-2

We have computed a smooth representation of the mean velocity $\langle
v\rangle $ and of the velocity dispersion $\sigma$ of the measurements
(cf.\ Merritt et al.\ \cite{merritt}). This was done by adaptive
kernel smoothing. For each star, we created a bin containing its 200
nearest neighbors, and computed the mean velocity $\langle v \rangle$
and velocity dispersion $\sigma$ for the bin, using a maximum
likelihood method with a Gaussian kernel centred on the star to
correct for the individual measurement errors (see Paper III).  This
procedure correlates the values at different points, but it produces
smooth maps of $\langle v \rangle$ and $\sigma$ which bring out the
main features of the line-of-sight kinematics of $\omega$ Cen. They
are shown in Figure~\ref{rv_F:v_field}. The mean velocity field is
very regular, and peaks at about 6 \kms\ at $8'$, beyond which it
decreases.  The contours of constant velocity dispersion are
elongated, and range from 15 \kms\ in the centre to 8 \kms\ at
10$'$. The decline continues to the edge of the field, where we
measure about 6 \kms\ at 25$'$.

We emphasize that the kinematic maps shown in
Figure~\ref{rv_F:v_field} have {\em not} been corrected for the
perspective rotation caused by the significant space motion of the
cluster and the relatively large field over which we have kinematic
measurements. We do this in Paper III, and show there that the
resulting maps remain smooth, but that the position angles of the zero
velocity curve and the orientation of the elongated contours of
constant velocity dispersion change significantly.

\section{Conclusions}
\label{rv_S:concl}

We have presented radial velocities obtained with {\tt ARGUS} for 2134
stars in the field of the globular cluster $\omega$ Centauri brigher
than photographic $B$-magnitude 16.5. We have shown that the standard
errors provided by the data reduction pipeline need to be multiplied
by 0.8. The externally-estimated median errors are below 2 \kms\ for
the stars brighter than $B$-magnitude of 15, which constitute the bulk
of the sample.  Based on these measurements, we conclude that 1589
stars are members of $\omega$ Centauri. The velocities provide clear
evidence for rotation with an amplitude of about 6 \kms. The velocity
dispersion is 15 \kms\ in the inner few arcmin, and decreases
monotonically to about 6 \kms\ at a radius of about $25'$.  We use
these velocities in a companion paper (Paper III) to correct the
proper motions from Paper I for the remaining overall (solid-body)
rotation, and then model the internal dynamics of $\omega$ Cen, and
determine its distance.

\begin{acknowledgements}
The authors thank Michael Perryman and an anonymous referee for
constructive comments on an earlier version of the manuscript, and the
Leids Kerkhoven-Bosscha Fonds, the Netherlands Organization for
Scientific Research (NWO) and Sterrewacht Leiden for travel support.
\end{acknowledgements}

\appendix

\section{Additional measurements}

We observed an additional 339 stars not discussed in the main
text. These consist of (i) 87 stars with $B-V<0.4$, and (ii) 252 stars
without colour information.\footnote{The number of stars with {\tt
ARGUS} radial velocity measurements quoted in Paper I did not
distinguish between our main sample of 1966, and the 339 discussed
here.}

The first group contains many stars whose colours put them on the
horizontal branch of $\omega$ Cen. About half of the velocities
provided by {\tt FXCOR} turned out to lie in the range covered by
members of $\omega$ Cen, while the others all appeared to lie in a
small interval near 20 \kms, which resulted in an anomalous peak in
the field star histogram of Fig.~\ref{rv_F:hist}.  As a result, we
suspect that many, if not all, of these horizontal-branch stars are in
fact cluster members, and that {\tt FXCOR} provided an erroneous value
for their radial velocity. For this reason, we do not trust the values
of the horizontal-branch `members' either, and decided to remove all
stars with $B-V<0.4$ from our main sample.  We list these stars and
their positions in Table~\ref{rv_T:extra}, together with our nominal
measured $\langle v \rangle$, the standard error $\sigma_{\langle v
\rangle}$ (0.8 times the standard error provided by {\tt FXCOR}, see
\S~2.6), the number of measurements, and the photographic $B$
magnitude.

Most of the stars without colour information were in the initial
preliminary proper motion catalog, but not in the final proper motion
catalog (see \S~2.2) which formed the basis of Table 5 in Paper I.
The bulk of the stars is expected to have $B-V \geq 0.4$, and the
resulting measurements should be reliable. As we do not know for
certain which of these stars are blue, and hence unreliable, we
decided to exclude the entire set from the analysis in the main
paper. They are all listed in Table~\ref{rv_T:extra}.

\begin{table}
\caption[]{Summary of results, extract only. LID: Leiden
identification number from Paper I; $\alpha$ and $\delta$ are right
ascension and declination in decimal degrees; mean measured
heliocentric radial velocity $\langle v \rangle$ and the standard
error $\sigma_{\langle v \rangle}$ (0.8 times the standard error
provided by {\tt FXCOR}), in \kms; $n$: number of individual
measurements; $B$: photographic magnitude in Mag.}
\label{rv_T:extra}
\begin{center}
\begin{tabular}{cccrccc}\hline\hline\noalign{\smallskip}
LID  & $\alpha$   & $\delta$
          & $\langle v\rangle$ & $\sigma_{\langle v \rangle}$ &$n$ &$B$ \\
\noalign{\smallskip}\hline\noalign{\smallskip}
 00002$\!\!$ &200.90579$\!\!$ & -47.15459$\!\!$ & -15.6$\!\!$
             &   5.2$\!\!$ &  $\!\!$1$\!\!$ &14.65 \\
 00004$\!\!$ &201.03029$\!\!$ & -47.15715$\!\!$ &   -42.6$\!\!$
             &   4.0$\!\!$ &  $\!\!$1$\!\!$ &15.56 \\
 00020$\!\!$ &202.16182$\!\!$ & -47.70901$\!\!$ &   -51.5$\!\!$
             &   1.9$\!\!$ &  $\!\!$1$\!\!$ &13.20 \\
 01002$\!\!$ &200.95038$\!\!$ & -47.16522$\!\!$ &     5.4$\!\!$
             &   2.0$\!\!$ &  $\!\!$1$\!\!$ &15.76 \\
 01004$\!\!$ &201.11462$\!\!$ & -47.16870$\!\!$ &   -74.1$\!\!$
             &   3.3$\!\!$ &  $\!\!$1$\!\!$ &16.34 \\
 01005$\!\!$ &201.12079$\!\!$ & -47.16557$\!\!$ &    51.3$\!\!$
             &   2.8$\!\!$ &  $\!\!$1$\!\!$ &15.46 \\
\noalign{\smallskip}\hline
\end{tabular}
\end{center}
\end{table}

\section{Mean and Dispersion}

Given a sample of $N$ radial velocities $v_i$ ($i=1,\dots,N$) with
corresponding measurement errors $\sigma_i$, we calculate the mean
$\langle v\rangle$ of this sample as
\begin{equation}
  \label{eq:meandisp_estimator}
  \langle v \rangle = \frac1S \sum_{i=1}^N w_i v_i,
\end{equation}
and the dispersion $s$ as
\begin{equation}
  s^2 = {1\over b^2(N)} \, {1\over S} \sum_{i=1}^N w_i
         \left(v_i-\langle v \rangle\right)^2,
\end{equation}
with weights $w_i=1/\sigma_i^2$ and $S=\sum_{i=1}^N w_i$. The
factor
\begin{equation}
  N\,b^2(N) = {2\Gamma^2\left(\frac{N}{2}\right) \over
              \Gamma^2\left(\frac{N-1}{2}\right)}
              \approx N - \frac32,
\end{equation}
with $\Gamma(z)$ the gamma function, makes $s$ an unbiased
estimator of the dispersion\footnote{If $N\,b^2(N)$ is replaced by
$N-1$, we obtain the well-known unbiased estimator of the
variance.}. Assuming that the velocities are Gaussian distributed,
the uncertainties $\sigma_{\langle v \rangle}$ of the mean, and
$\sigma_s$ of the dispersion, are given by
\begin{equation}
  \label{eq:meandisp_uncertainty}
  \sigma_{\langle v \rangle} = {1\over \sqrt{S}}
\end{equation}
and
\begin{equation}
  \sigma^2_s = \sigma^2_{\langle v \rangle}
  \left[{N-1 \over N\,b^2(N)} - 1\right] \approx
  {N \sigma^2_{\langle v \rangle} \over 2N-3}.
\end{equation}


\begin{thebibliography}{}
%
%
\def\aa{{A\&A} }
\def\aas{{A\&AS} }
\def\aj{{AJ} }
\def\apj{{ApJ} }
\def\apjs{{ApJS} }
\def\araa{{ARA\&A} }
\def\mnras{{MNRAS} }
\def\pasp{{PASP} }
\def\baas{{BAAS} }


\bibitem[2002]{and02}
Anderson J., 2002, in F.\ van Leeuwen, J.D.\ Hughes, G.\ Piotto (eds),
$\omega$ Centauri, A Unique Window into Astrophysics. ASP Conf.\ Ser.\
265, 87

\bibitem[1902]{bailey}
Bailey S.I., 1902, Harvard Annals 38

\bibitem[2004]{bpa+04}
Bedin L.R., Piotto G., Anderson J., Cassisi S., King I.R., Momany Y.,
Carraro G., 2004, ApJL 605, 125

\bibitem[1967]{dw67}
Dickens R.J., Woolley R.v.d.R., 1967, Royal Obs.\ Bull.\ No 128

\bibitem[1962]{fd62}
Fehrenbach C.H., Duflot M., 1962, ESO Comm.\ No.2

\bibitem[2004]{fsp+04}
Ferraro F.R., Sollima A., Pancino E., Bellazzini M., Straniero O., Origlia L.,
Cool, A.M., 2004, ApJ 603, L81

\bibitem[1983]{geyer2}
Geyer E.H., Hopp U., Nelles B., 1983, A\&A 125, 359

\bibitem[1965]{harding}
Harding G.A., 1965, Royal Obs.\ Bull.\ No 99

\bibitem[2004]{hwl+04}
Hughes J.A., Wallerstein G., van Leeuwen F., Hilker M., 2004, AJ 127, 980

\bibitem[1988]{ingerson}
Ingerson T., 1988, In: Barden S.C.\ (ed.) Fiber Optics in Astronomy.
ASP Conf.\ Ser.\ 3, 99

\bibitem[1966]{king66}
King I.R., 1966, AJ 71, 64

\bibitem[1999]{ljs+99}
Lee Y.-W., Joo J.-M., Sohn Y.-J., Rey S.-C., Lee H.-C., Walker A.R.,
1999, Nature 402, 55

\bibitem[2000]{llrfz00}
van Leeuwen F., Le Poole R.S., Reijns R.A., Freeman K.C., de Zeeuw P.T., 2000
A\&A 360, 472 (Paper I)

\bibitem[2002]{vlhp02}
van Leeuwen F., Hughes J., Piotto G., 2002, $\omega$ Centauri: A
Unique Window into Astrophysics. ASP Conf.\ Ser.\ 265

\bibitem[2002]{lub02}
Lub J., 2002, in F.\ van Leeuwen, J.D.\ Hughes, G.\ Piotto (eds),
$\omega$ Centauri, A Unique Window into Astrophysics. ASP Conf.\ Ser.\
265, 95

\bibitem[1990]{lutz}
Lutz T.E., Ingerson T., Schumacher G., Smith D., 1990, PASP 102, 1208

\bibitem[1938]{martin}
Martin W.C., 1938, Annalen Sterrewacht Leiden, deel XVII

\bibitem[1996]{mayor96}
Mayor M., Duquennoy A., Alimenti A., Andersen J., Nordstr\"om B., 1996, In:
Milone G., Mermilliod J.-C. (eds.)  The Origins, Evolution, and Destinies of
Binary Stars in Clusters. ASP Conf.\ Ser.\ 90, 190

\bibitem[1997]{mm_data}
Mayor M., Meylan G., Udry S., Duquennoy A., Andersen J., Nordstr\"om B.,
Imbert M., Maurice E., Pr\'evot L., Ardeberg A., Lindgren H., 1997,
AJ 114, 1087

\bibitem[1997]{merritt}
Merritt D., Meylan G., Mayor M., 1997, AJ 114, 1074

\bibitem[1995]{meylan}
Meylan G., Mayor M., Duquennoy A., Dubath P., 1995, A\&A 303, 761 

\bibitem[1995]{nordca}
Norris J.E., Da Costa G.S., 1995, ApJ 447, 680

\bibitem[1996]{norris96}
Norris J.E., Freeman K.C., Mighell K.J., 1996, ApJ 462, 241

\bibitem[2000]{pfb+00}
Pancino E., Ferraro F.~R., Bellazzini M., Piotto G., Zoccali M., 2000,
ApJL 534, 83

\bibitem[2002]{pan02}
Pancino E., 2002, in F.\ van Leeuwen, J.D.\ Hughes, G.\ Piotto (eds),
$\omega$ Centauri, A Unique Window into Astrophysics. ASP Conf.\ Ser.\
265, 313

\bibitem[1891]{pickering}
Pickering E.C., 1891, Harvard Annals, 26

\bibitem[2005]{pvb+05} 
Piotto G., Villanova S., Bedin L.R., Gratton R., Cassisi S., Momany
Y., Recio-Blanco A., Lucatello S., Anderson J., King I.R.,
Pietrinferni A., Carraro G., 2005, ApJ 621, 777\looseness=-2

\bibitem[1987]{spitzer}
Spitzer L.\ jr., 1987, In: Dynamical Evolution of Globular Clusters, Princeton
University Press, p.\ 191

\bibitem[1996]{sk96}
Suntzeff N.B., Kraft R.P., 1996, AJ 111, 1913

\bibitem[1986]{tody}
Tody, D., 1986 In: Crawford D.L.\ (ed.) The IRAF Data Reduction and Analysis
System, Instrumentation in Astronomy VI, Proc.\ SPIE, 627, 733

\bibitem[1979]{tonry}
Tonry J., Davis M., 1979, AJ 84, 1511

\bibitem[1995]{tkd95}
Trager S.G., King I.R., Djorgovski S.C., 1995, AJ 109, 218

\bibitem[1991]{tyson}
Tyson N.D., Rich R.M., 1991, ApJ 367, 547

\bibitem[2005]{vbv+05}
van de Van G., van den Bosch R.C.E., Verolme E.K., de Zeeuw P.T., 2005, 
AA, submitted (Paper III)

\bibitem[1966]{woolley}
Woolley R.v.d.R., 1966, Royal.\ Obs.\ Annals, No 2

\end{thebibliography}
\end{document}